\documentclass{actasen}

\begin{document}

\setcounter{page}{67}

\Volume{2015}{56}


\runheading{Tatsuyuki~TAKATSUKA }%

\title{Massive Neutron Stars with Hadron-Quark Transition Core
---phenomenological approach by \lq\lq3-window model\rq\rq ---}

\footnotetext{$^{\bigtriangleup}$ takatuka@iwate-u.ac.jp}

Science B. V. All rights reserved. 

\noindent PII:

\enauthor{Tatsuyuki TAKATSUKA$^{\dag}~\!^{\bigtriangleup}$,Kota MASUDA$^{\star}$ and Tetsuo HATSUDA$^{\dag}~\!^{\star\star}$ }
{${\dag}$Theoretical Research Division, Nishina Center, RIKEN, Wako 351-0198, Japan\\
${\star}$Department of Physics, The University of Tokyo, Tokyo 113-0033, Japan\\
${\star\star}$Kavli IPMU, The University of Tokyo, Kashiwa 277-8583, Japan}

\abstract{By a new approach introducing a \lq\lq 3-window model\rq\rq and constructing phenomenologically an equation of state for the hadron-quark (HQ) transition
 region,  possible maximum mass of neutron stars (NSs) is discussed.  
It is found that neutron stars (NSs) with HQ transition core are able to have a mass exceeding $2M_{\odot}$, consistent with massive NSs recently observed.}


\maketitle

\section{Introduction}

 Recent observations of a two-solar-mass neutron star ($2M_{\odot}$-NS) \rf{1, 2}  present a challenging problem how to accommodate a very stiff equation of state (EOS) responsible for such massive NSs.  
This problem is serious since every new exotic phase (e.g., meson condensation, 
hyperon ($Y$) mixing, quark (Q) matter) leads to a softening of EOS against the requirement 
from observations.  For example, the $Y$-mixing softens the EOS dramatically and thereby the NS 
maximum mass ($M_{\max}$) goes down to (1.1-1.2)$M_{\odot}$ (\lq\lq Hyperon Crisis\rq\rq),
 clearly contradicting even the \lq\lq minimal mass\rq\rq 1.44$M_{\odot}$ observed for PSR1913+16 \rf{3, 4 }. 

The aim of this paper is to discuss how NSs could be massive, by introducing a quark degrees of freedom.  
Our strategy for the approach is to divide the EOS into three density ($\rho$) regime i.e., pure hadronic matter EOS (H-EOS, $\rho\leq\rho_H$), hadron-quark transition matter EOS (HQ-EOS, $\rho_H\leq\rho\leq\rho_Q$) and pure quark matter EOS (Q-EOS, $\rho\geq\rho_Q$), which we call \lq\lq3-window model\rq\rq \rf{5-9}.  
This is motivated by the following considerations: Pure hadron-matter EOS calculated from point-like hadrons 
plus their interactions looses the validity with increasing $\rho$, primarily because baryons have a finite size composed of quarks (and gluons) and deconfinement effects come into play.  
Also pure Q-matter EOS gets uncertain with decreasing $\rho$ because strong correlations among quarks would develop and confinement effects would participate.  
Our basic idea is to supplement the very poorly known HQ-EOS relevant to a confinement-deconfinement transition, by sandwiching it in between the relatively certain H-EOS at lower density side and Q-EOS at higher density side.
We stress that our new approach to the H-Q transition is not restricted by a conventional Gibbs or Maxwell condition which necessarily leads to a softening of EOS.
 
\begin{figure}[thb]
\centering
{\includegraphics[angle=0,width=8cm]{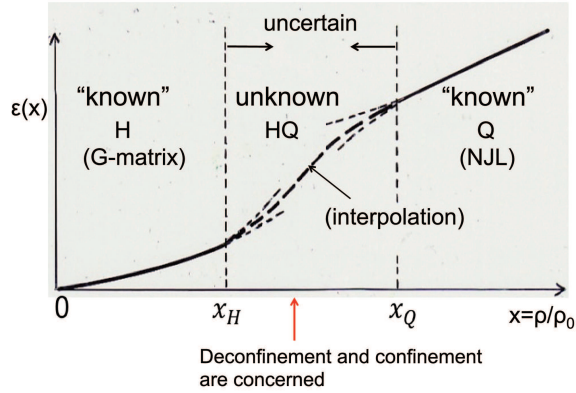}}
\caption{Schematic illustration of \lq\lq3-window model\rq\rq.  
Energy density $\epsilon$ versus $\rho$.}
\end{figure}

\section{Construction of EOS}
We use the H-EOS from a G-matrix-based effective interaction approach applied to neutron-star matter 
composed of $N(\equiv n$, $p$), $Y(\equiv\Lambda$, $\Sigma^-$) and leptons \{$e^-$, $\mu^-$\} being 
in $\beta$-equilibrium and charge neutrality\rf{4}.  
The H-EOS is designed to satisfy the saturation properties of symmetric nuclear matter and gives nuclear 
incompressibility of $\kappa=250$ MeV consistent with experiments.  
As for the Q-EOS, we use that from an effective theory of QCD, namely, the (2+1)-flavor  NJL model with 
the Hatsuda-Kunihiro parameter set, where Q-matter composed of \{$u$, $d$, $s$\} quarks and leptons 
\{$e^-$, $\mu^-$\} are in a charge neutral and $\beta$-equilibrated system\rf{6, 7}.  
In the model Lagrangian, we include a phenomenological vector-type interaction with the strength $g_v$ 
which leads to an universal flavor-independent repulsion among quarks, contributing to a stiffening of the Q-EOS.  

Concerning the HQ-EOS, we obtain it by a phenomenological interpolation taking the interpolation function $\epsilon_{HQ}(\rho)$ of a form of 5-th degree polynomial with 6-parameters.  
The 6-parameters are determined by the 6-conditions that the energy density $\epsilon(\rho)$, 
the pressure $p(\rho)(=\rho^2\partial(\epsilon/\rho)/\partial\rho)$ and the sound velocity $v_s(\rho)(=c\sqrt{\partial p/\partial\epsilon})$ are coincide with each other at two boundaries ($\rho=\rho_H$ and $\rho=\rho_Q$).  
The $\epsilon_{HQ}(\rho)$ is restricted by the conditions, $p\geq0$, $\partial p/\partial\rho\geq0$ and $v_s\leq c$.

\begin{table}[h]
\centering
\caption{Some results of maximum mass ($M_{max}$) neutron stars.}
\fns\tabcolsep 2.2mm
\begin{tabular}{ccccc}
\hline
 & $g_v/G_S=0$ & & $g_v/G_S=0.5$ & \\
\hline
($x_H$, $x_Q$) & (1.5, 11) & (1.5, 8.5) & (1.5, 11) & (2, 11) \\
\hline
$M_{max}/M_{\odot}$ & 1.79 & 2.36 & 2.21 & 2.20 \\
$R$(km) & 10.2 & 11.4 & 10.8 & 10.4 \\
$\rho_c/\rho_0$ & 7.25 & 5.32 & 6.04 & 6.33 \\
\hline
\end{tabular}
\begin{minipage}[b]{14cm}{\footnotesize
\vspace*{1mm}\hspace*{0.3cm}$g_v(G_S)$ denotes the strength of a vector (scalar) interaction.  $R(\rho_c)$ is the radius (central density) of NSs.}
\end{minipage} \end{table}

\section{Results and concluding remarks}

We have studied how massive could be the hybrid stars with quark degrees of freedom, 
by phenomenologically constructing the EOS with hadron-quark transition in a new approach 
characterized by \lq\lq3-window model\rq\rq.  
Some numerical results are shown in Table 1 where $R(\rho_c)$ is the radius (central density).  
It is found that NSs with a HQ transition core have a potentiality to generate 
massive NSs, e.g., with $M=(2.2-2.4)M_{\odot}$, as far as a quark degrees of freedom sets in 
at rather low-density ($(1.5-2)\rho_0$,with $\rho_0$=0.17/fm$^3$ being the nuclear density) due to the percolation of quarks in hadronic matter\rf{10} 
and the EOS of quark matter is stiff.  
The results from the present approach confirm those of our preceding works performed from 
a HQ crossover picture \rf{6, 7}.  

Finally, we want to remark that the possible candidates to resolve the "Hyperon Crisis" problem are 
the \lq\lq universal 3-body force\rq\rq when a purely hadronic scheme is taken, as shown previously\rf{11}, 
and the HQ transition occurring in NS cores when \{hadron+quark\} scheme is considered.

\end{document}